\documentclass{article}
\usepackage{spconf,amsmath,graphicx}
\usepackage{color}
\usepackage{fancyhdr}
\usepackage{graphicx}
\usepackage{amsthm}
\usepackage{amsmath}
\usepackage{amssymb}
\usepackage{longtable}
\usepackage{mathrsfs}
\usepackage{enumerate}
\usepackage{indentfirst}
\usepackage[colorlinks,linkcolor=blue]{hyperref}
\newcommand{\tabincell}[2]{\begin{tabular}{@{}#1@{}}#2\end{tabular}} 

\title{Musical Instrument Playing Technique Detection Based on FCN: Using Chinese Bowed-Stringed Instrument as an Example}
\name{Zehao Wang$^*$$^1$, Jingru Li\sthanks{Equal Contribution.}$^1$, Xiaoou Chen$^2$, Zijin Li$^3$, Shicheng Zhang$^4$, Baoqiang Han$^3$, Deshun Yang$^2$}
\address{$^1$Peking University, Beijing, China  \quad
$^2$Wangxuan Institute of Computer Technology, PKU, Beijing, China\\
$^3$China Conservatory of Music, Beijing, China \quad
$^4$University of Illinois at Urbana-Champaign, Illinois, USA \\
\{water45wzh, li\_jingru, chenxiaoou, yangdeshun\}@pku.edu.cn, zijin.li@mcgill.ca, \\sz18@illinois.edu, hundel@126.com}
\begin{document}
	%
	\maketitle
	\begin{abstract}
		Unlike melody extraction and other aspects of music transcription, research on playing technique detection is still in its early stages. Compared to existing work mostly focused on playing technique detection for individual single notes, we propose a general end-to-end method based on Sound Event Detection by FCN for musical instrument playing technique detection. In our case, we choose Erhu, a well-known Chinese bowed-stringed instrument, to experiment with our method. Because of the limitation of FCN, we present an algorithm to detect on variable length audio. The effectiveness of the proposed framework is tested on a new dataset, its categorization of techniques is similar to our training dataset. The highest accuracy of our 3 experiments on the new test set is 87.31\%.  Furthermore, we also evaluate the performance of the proposed framework on 10 real-world studio music (produced by midi) and 7 real-world recording samples to address the ability of generalization on our model.

	\end{abstract}
	\begin{keywords}
		Music Information Retrieval, Playing Technique Detection, Sound Event Detection
	\end{keywords}
	\section{Introduction}
	\label{sec:intro}
	
	In addition to the recognition and extraction of music notions such as pitches and chords, the detection of musical instrument playing techniques also plays an important role in music transcription. In general, a note-by-note transcription of the pitches and the playing techniques associated with each note is demanded. In a sequence of melodic notes, playing techniques such as slide and vibrato determine how the notes are performed and sounded. Articulation on any form of representation of music, including western staff notation and Chinese numeric representation\footnote{Called \textit{Jianpu} in Chinese}, is highly instructive for performers.  This can be illustrated in \textbf{Fig. 1} through the articulation symbols on the guitar tab and \text{Jianpu} used by Erhu\footnote{A famous Chinese bowed-stringed instrument like the violin} and Guqin\footnote{A famous Chinese stringed instrument like a harp lying flat}. 
	
	Playing technique detection from audio recording can help us to make automatic music transcription more accurate. Nowadays there are useful methods for music transcription, such as through F0\cite{canadas2010multiple}, CQT\cite{argenti2010automatic}...Perhaps it is more suitable to call them as melodic extraction. Currently, the mentioned methods are designed for solo music instrument note extraction, but if we want to expand it to all types of musical instruments, the notations for playing technique are necessary. We can take the articulation, trill, as an example. Since it contains large number of different notes in a single occurrence, the time value of each note is significantly short. The threshold of post-processing or other parameters in the melody extraction algorithm is not suitable for this technique, therefore some pitches may be ignored. Consequently it might lose some important information for accurate music transcription. 
	
	The realization of playing technique detection from audio recording may also exerts transformative force in industrial and educational applications. Applications, such as playing-technique-detection-based audio content analysis and music assisted teaching, including musical instrument playing or Orchestration course in music composition, may be good examples for illustration. 
	
	Unlike melody extraction and other aspects of music transcription, research on playing technique detection is still in its early stages. The difficulties of this task are mainly about data acquisition and data labeling. Currently, the research on musical instrument playing technique detection involves guitar\cite{chen2015electric}\cite{su2014sparse}, piano\cite{liang2019piano}\cite{liangtowards}, drum\cite{wu2016drum}...Among all the musical instruments, the playing techniques of the bowed-stringed are difficult to detect and manually label even for performers because of its variety and uncertainty of onset, so the number of research in this field is not many.
	
	In this paper, we propose a general end-to-end method based on Sound Event Detection by FCN\cite{long2015fully}(Fully Convolutional Networks) through a deep learning method for our musical instrument playing technique detection task. Considering the data acquisition and variety of playing techniques, we choose Erhu, a well-known Chinese bowed-stringed instrument, to experiment with our method. The contribution of this work is listed below. First, we compile an open dataset that contains about 30 Erhu playing techniques in 11 categories, covering almost all the tones in the range of Erhu (\textbf{Section 3}). We have released experiment's Python code\footnote{\href{https://github.com/water45wzh/MIPTD_Erhu}{https://github.com/water45wzh/MIPTD\_Erhu}}and the full datasets with detailed information and demos\footnote{\href{https://water45wzh.github.io/MIPTD_Erhu}{https://water45wzh.github.io/MIPTD\_Erhu}} online. Second, we propose a method having a low dependency on the sound characteristics of the tested instruments, ensuring its transferability to other audio clips recorded from different musical instruments. Third, we have conjectured several ways for its industrial application including boosting efficiency on the traditional music transcription task and realizing MIR-based music education pedagogy.

	\section{Related Work}
	\label{sec:related}
	Currently, the research on Musical Instrument Playing Technique Detection (MIPTD) mainly focuses on playing techniques classification of the single block of musical note. Through the ages, there are both traditional signal processing methods and statistical/deep learning methods for this task. Chen \textit{et al.}\cite{chen2015electric} attempted to extend this research to guitar solo recordings, they considered the playing techniques as time sequence pattern recognition problem and developed a two-stage framework for detecting five fundamental playing techniques used by electric guitars. This work is validated on 42 electric guitar solo tracks without accompaniment, and also discussed how to apply the framework to the transcription of real-world electric guitar solo with accompaniment.
	
	Wu \textit{et al.}\cite{wu2016drum} studied the drum playing technique detection in polyphonic mixtures. They focused on 4 rudimentary techniques: strike, buzz roll, flam, and drag. This paper has discussed about the characteristic and challenge of this task, compared with different sets of features, like features extracted from NMF-based activation functions, as well as baseline spectral features.  And they also considered it's difficult to detect playing techniques from polyphonic music.
	
	Liang \textit{et al.}\cite{liangtowards} thought that detecting playing techniques from audio signal of musical performance is a special aspect of automatic music transcription. They have studied deeply on piano sustain pedal detection, and given some useful datasets and frameworks to analyze the existence of pedal. The latest work of Liang\cite{liang2019piano} is about a joint model by CNN on detecting the pedal onset and segment with data fusion method to make piano sustain pedal detection more accurate.

	\section{Datasets and Erhu Playing Techniques}
	\label{sec:datasets}
	
	Two datasets are used in this paper. The first one is based on Chinese traditional musical instruments dataset DCMI \cite{lidcmi}. This dataset contains solo instrument playing technique recordings on 82 different Chinese traditional musical instruments with almost all types of playing techniques of the instruments. We choose a subset of Erhu's playing techniques from DCMI and edit it to 927 audio clips of single playing techniques in 11 categories covering almost all the tones in the range of Erhu, their time length ranges from about 0.2 to 2 seconds, in total 18.45 minutes. We called these audio clips \textit{short clips}. For the brevity of this paper, details about the dataset and the involved Erhu playing techniques can be found on the website. The second one is similar to the first, in total 7.36 minutes of 326 short clips, with changes of variable on performer, instrument, and recording environment. This variation is considered for generalization performance on different data distribution. 
	
	\begin{figure}
		\centering
		\includegraphics[width=8cm]{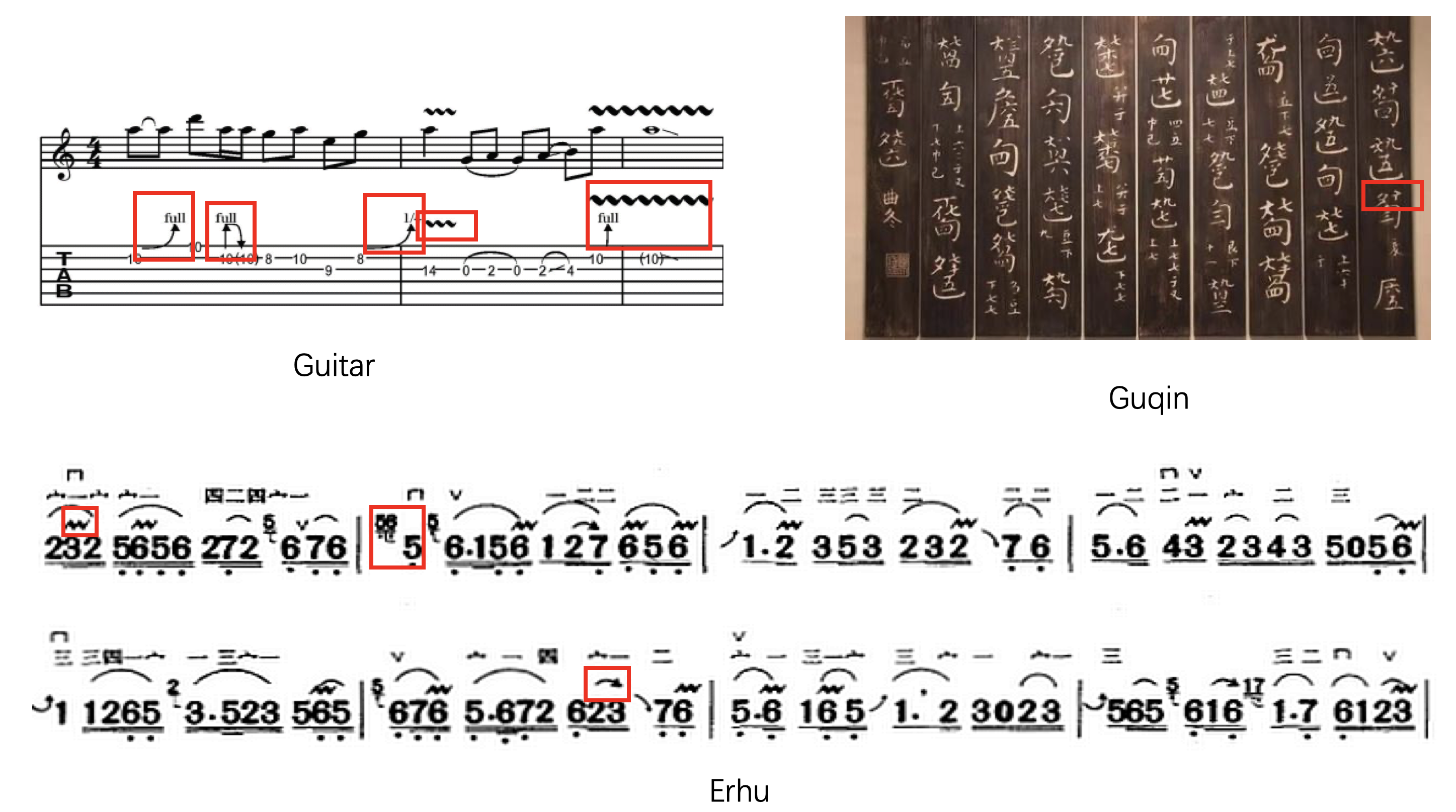}
		\caption{Playing techniques of Guitar, Erhu, Guqin}
		\label{fig:tech}
	\end{figure}
	
	On both training and test, we randomly generate some 10 seconds long audio segments from these two datasets mentioned in last paragraph, we called these audio segments \textit{long segments}. There is a basic assumption that the techniques of Erhu are not overlapped. The rationale of this assumption is due to the limitation of monophonic instruments. During the generation process, we execute a check operation to avoid the repetition of long segments sequence. To avoid the uncertainty of the length of the generated long segment, we directly trim the excess for more than 10 seconds. To ensure the long segments sound more realistic, we make a 50 milliseconds cross-fade in each boundary of adjacent two short clips. Since the short clips are almost recorded in D major scale, the random generated long segments sound closed to the real-world music. 
	
	We also label the timestamp of corresponding playing techniques during the long segments generation process. The format of the label consists of event tags recorded with a frame length of 0.05 second. Due to the impossibility of playing multiple techniques simultaneously on a solo instrument, the above operation is reasonable.
	
	Furthermore, we have 10 real-world music produced by one of the best digital instrument library, \textit{Silk} from the famous company \textit{EastWest}\footnote{EastWest Sounds(\href{http://www.soundsonline.com/}{http://www.soundsonline.com/})}, and 7 real-world recorded samples of Erhu solo by famous artists and corresponding manual event based label. The detailed involving audio's information and demo for their experiments can be found online\footnote{\href{https://water45wzh.github.io/MIPTD_Erhu}{https://water45wzh.github.io/MIPTD\_Erhu}}.
	
	The sample rate of all the audios in our datasets is 44.1kHz. 
	
	\begin{figure}
		\centering
		\includegraphics[width=7cm]{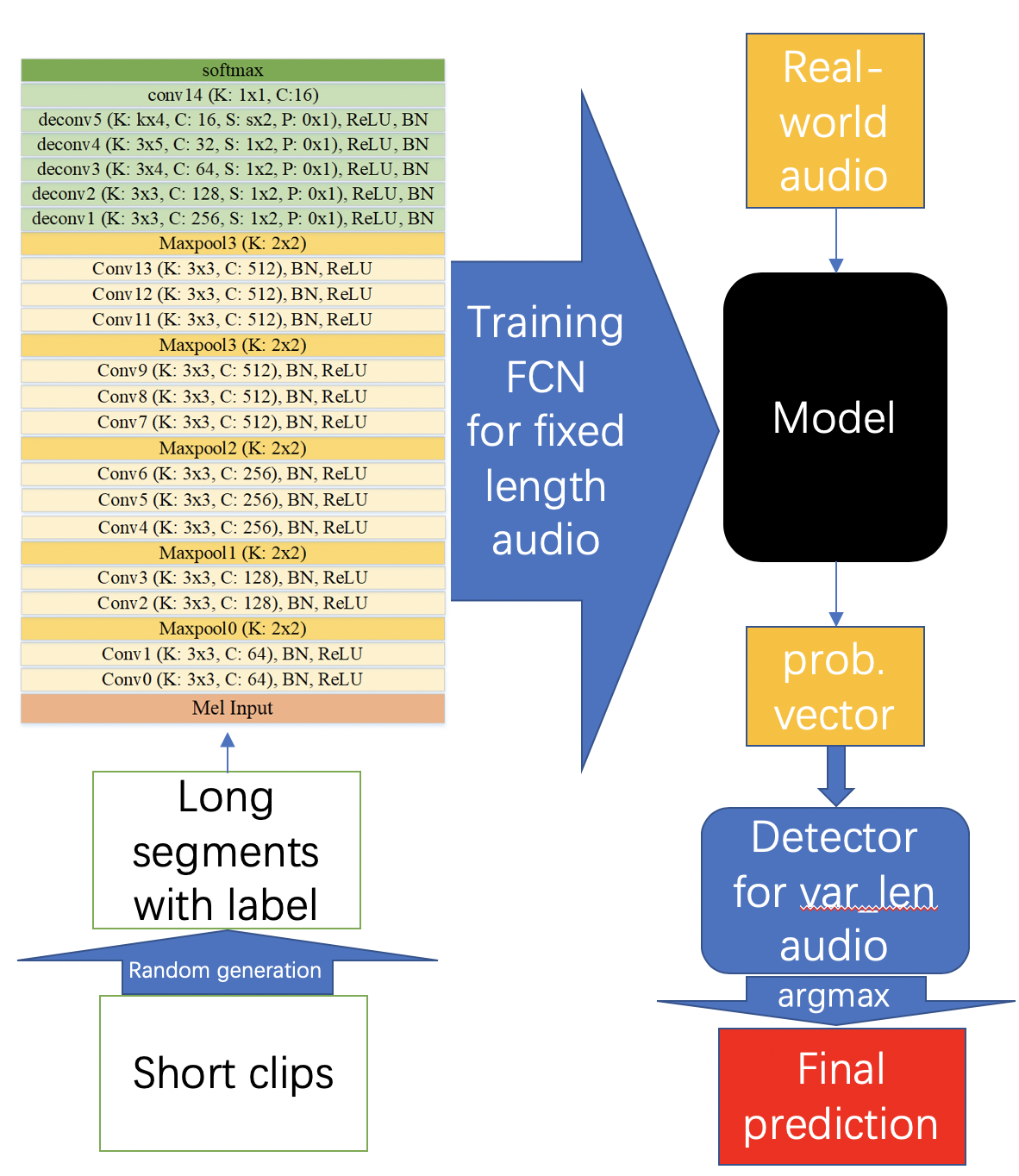}
		\caption{Framework of our work. In deconv5, 4 classes: $k=3, s=1$, 7 classes: $k=3, s=2$, 11 classes: $k=4, s=3$.}
		\label{fig:framework}
	\end{figure}
	
	\section{Proposed Method}
	\label{method}
	\subsection{Overview}
	
	Musical instrument playing technique detection (MIPTD) is essentially a task of sound event detection. We can think of each technique as a sound event and recognize the event from the sound event stream. Currently CNN is widely used in the field of sound event detection\cite{su2017weakly} frequently using spectrograms or mel-spectrograms as inputs. Because of the non-overlap feature of monophonic instrument, we can naturally treat it as the semantic segmentation in image segmentation tasks.
	
	Recent years, researchers proposed several approaches for the segmentation task. We choose Fully Convolutional Networks\cite{long2015fully}\cite{su2017weakly}, i.e. FCN for our task, which is a classical approach for image semantic segmentation. This approach constructs an end-to-end \textit{Fully Connected Convolutional} network, which is one of the state-of-the-art approaches for semantic segmentation task. In this paper, we have adapted it to our musical instrument playing technique detection task.
	
	The main limitation of FCN is that it can only be used for fixed length audio. To solve this problem, we make a framework to detect variable length audio. Details are elaborated in \textbf{4.3}. 
	
	The overview of the whole framework for this task is shown in \textbf{Fig. 2}.

	\subsection{Network Architecture}
	
	Our network architecture is also shown in Fig. 2. Its input is a mel-spectrogram and output is a $k$-dim probabilistic vector, $k$ is the number of classes(4, 7, 11).
	
	\subsection{Detection on Variable Length Audio}
	
	Our trained model from above neural network is specified for 10 seconds audio. For the detection on variable length audio, we use the fixed length(10s) of the model as the window length and set a suitable hop length(we choose 2s) to slide sequentially from the start to the end on the entire audio recording. For a fixed 0.05 second frame of the audio, if there are more than two predictions on this frame due to overlapped predictions, then we might compute the average of all the predicted probabilistic vectors of this frame as the final prediction for it,
	\begin{equation}
		\mathbf{p}_{final} = \frac{1}{|\mathbf{P}|}\sum_{\mathbf{p} \in \mathbf{P}}\mathbf{p}
	\end{equation}
	For an arbitrary frame, $\mathbf{p}_{final}$ is the final prediction, $\mathbf{P}$ is a set of all the predictions for this frame. 
	
	\section{Experiments}
	\label{experiments}
	\subsection{Experimental Setup}
	We divide each long segment in our generated datasets into some frames with a frame length of 0.05 second using mel-filters to extract 128bin mel-spectrogram with window length of 2048 points and hop length of 2205 samples. We use \textit{PyTorch} for our implementations. 
	
	We initiate 3 experiments for different goals. Each experiment uses the first dataset or a subset of it to generate some 10s long segments for training. The second is also using the dataset with its subset but to generate several 10 seconds long segments for testing the generalization performance on high-quality midi music and recorded samples of our model.
	
	4 classes: We choose 3 common techniques (slide, staccato, trill) from 11 categories and 1 class for a subset of detache and other techniques. Training set has 2000 long segments, Test set has 1000.
	
	7 classes: We consider some subdivisions in all 11 categories, then choose 6 common subdivisional techniques (trill, slide\_up, slide\_down, staccato, trill\_up\_short, slide\_legato) and 1 class for a subset of detache and other techniques. Training: 4000, Test: 2000.
	
	11 classes: This experiment uses the full version of the first dataset, all of which are identical to the 11 categories. Training: 4000, Test: 2000.
	
	The reason of doing 3 experiments is because the 4 classes experiment is a simplified version of 11 classes experiment, in order to pick out some common techniques for common music. The purpose of the 7 classes experiments is to make a precise detection for the subdivisions of these techniques. A possible future work is to make a two-level system to realize the precise detection for the subdivisions.
	\begin{figure}
		\centering
		\includegraphics[width=8cm]{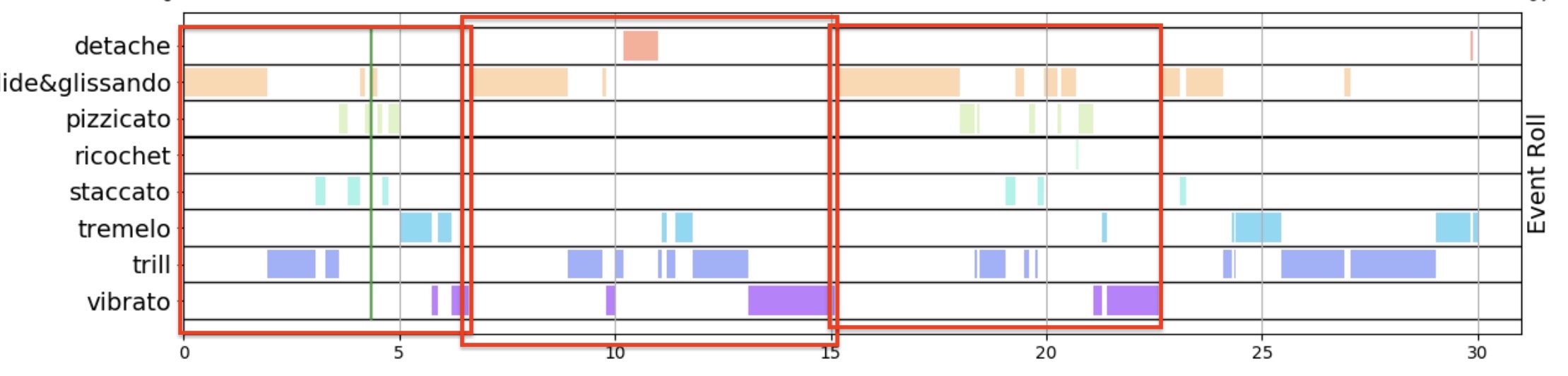}
		\caption{Visualization for a studio music}
		\label{fig:sbdys11}
	\end{figure}
	\subsection{Results on the test set}
	The results of 3 experiments for our test set from the second dataset are shown in the following table,
	
	\begin{tabular}{|l|c|c|c|}
		\hline
		& 4 classes & 7 classes & 11 classes \\
		\hline
		\tabincell{l}{\textbf{average} \\ \textbf{accuracy}} & \textbf{87.31\%} & \textbf{67.94\%} & \textbf{48.26\%} \\
		\hline
	\end{tabular}
	\\
	\\
	Because of short time value of some techniques (only 0.15-0.20 second, 3-4 frames) and end-to-end feature in our proposed model, we decide not to implement post-process for the output prediction. The accuracy of one segment of these test data is calculated strictly by the following formula,
	\begin{equation}
		accuracy\_of\_one\_segment\ =\ \frac{1}{n}\sum_{i=1}^{n} \mathbf{I}(p_i = l_i).
	\end{equation}
	
	The input audio has $n$ frames. Its ground-truth label is $\mathcal{L} = (l_1,l_2,...,l_n)$ and the output prediction by our model is $\mathcal{P} = (p_1,p_2,...,p_n)$. $\mathbf{I}(*)$ is an \textit{indicator function}. 
	
	\subsection{Results on real-world music}
For brevity of this paper, we will demonstrate several results on real-world music, both studio music and recorded samples. We use \textit{sed\_vis}\footnote{Visualization toolbox for Sound Event Detection(\href{https://github.com/TUT-ARG/sed_vis}{https://github.com/TUT-ARG/sed\_vis})} to visualize our results. As shown in \textbf{Fig. 3}, here is an illustration for a studio music, which is called \textit{The Umbrella of Cherbourg}, there are 4 sentences in this piece, and the sequences of techniques used in the first three sentences are the same. And results of several recording samples are as follows.
\\
\\
	\noindent
\begin{tabular}{|l|c|c|c|c|}
	\hline
	music & time & \tabincell{l}{acc. of \\4 classes}  & \tabincell{l}{acc. of \\7 cl.} & \tabincell{l}{acc. of \\11 cl.}\\
	\hline
	\tabincell{l}{The Moon's\\Reflection\\on the Sec-\\ond Spring} & 6m04s &\textbf{40.90\%} &\textbf{32.26\%} &\textbf{17.06\%} \\
	\hline
	\tabincell{l}{Competing \\Horses-v\_1} & 1m46s &\textbf{28.84\%} &\textbf{44.50\%} &\textbf{12.39}\% \\
	\hline
\end{tabular}

\section{Conclusion}
In this paper, we have proposed an approach for Musical Instrument Playing Technique Detection using Erhu as an example. We took advantage of FCN to model a fixed length detector and then proposed an algorithm for variable length audio. Without any assumptions and requirements of data distribution, we have successfully trained an end-to-end model using long segments randomly generated from 927 playing technique audio clips. Because of its end-to-end feature and slack dataset requirements, this method has high transferability on dataset of other musical instruments. The best experiment result on the test set is 87.31\% . Our result on the real-world recording(44.50\%) from famous artists is not as good as the other data, because of possible error of manual labeling, the common issue of bowed-stringed instruments, and artistic expression from the different performers. One possible future work is to make a combined system to implement the complete automatic music transcription, which means not only the pitch extraction for notes but also the techniques detection for notations on notes. Although the current research is preliminary compared to the ones in pitch extraction, we hope it will raise more awareness on the importance of the Musical Instrument Playing Technique Detection(MIPTD).
	
	\section{Acknowledgements}
	Thanks to Zihan Han of Xidian University, he provided a lot of help for us to record the test data set.
	\vfill\pagebreak

	\bibliographystyle{IEEEbib}
	\bibliography{refs}
\end{document}